\documentclass[aps,prd,superscriptaddress,nofootinbib,tighten,preprint]{revtex4}
\include{header}
\usepackage{amsfonts}
\usepackage{amssymb}
\usepackage{amsmath}
\usepackage{graphicx}
\usepackage{hyperref}
\usepackage{placeins}
\usepackage{gensymb}
\usepackage{color}
\usepackage{times}
\usepackage{url,hyperref}
\usepackage{epstopdf} 
\def\slc#1{\setbox0=\hbox{$#1$}           
    \dimen0=\wd0                                 
    \setbox1=\hbox{/} \dimen1=\wd1               
    \ifdim\dimen0>\dimen1                        
       \rlap{\hbox to \dimen0{\hfil/\hfil}}      
       #1                                        
    \else                                        
       \rlap{\hbox to \dimen1{\hfil$#1$\hfil}}   
       /                                         
    \fi}

\begin{document}

\title{Prospects of indirect searches for dark matter annihilations in the earth with ICAL@INO}
\author{Deepak Tiwari}
\email{deepaktiwari@hri.res.in}
\affiliation{Harish-Chandra Research Institute (HBNI), Chhatnag Road, Jhunsi, Allahabad 211 019, India}

\author{Sandhya Choubey}
\email{sandhya@hri.res.in}
\affiliation{Harish-Chandra Research Institute (HBNI), Chhatnag Road, Jhunsi, Allahabad 211 019, India}
\affiliation{Department of Physics, School of
Engineering Sciences, KTH Royal Institute of Technology, AlbaNova
University Centre, 106 91 Stockholm, Sweden}

\author{Anushree Ghosh}
\email{anushree.ghosh@usm.cl}
\affiliation{Universidad Tecnica Federico Santa Maria - Departamento de Fisica Casilla 110-V, Avda. Espana 1680, Valparaiso, Chile}

\begin{abstract}
We study the prospects of detecting muon events at the upcoming Iron CALorimeter (ICAL) detector to be built at the proposed India-based Neutrino Observatory (INO) facility due to neutrinos arising out of
annihilation of Weakly Interactive Massive Particles (WIMP) in the centre of the earth. The atmospheric neutrinos coming from the direction of earth core presents an irreducible background. 
We consider 50kt $\times$ 10 years of ICAL running and WIMP masses between 10-100 GeV and present 90 \% C.L. exclusion sensitivity limits on $\sigma_{SI}$ which is the WIMP-nucleon Spin Independent (SI) 
interaction cross-section. The expected sensitivity limits calculated for ICAL for the WIMP annihilation in the earth are more stringent than the limits obtained by any other indirect detection experiment. 
For a WIMP mass of ~$52.14 \textup{ GeV}$, where the signal fluxes are enhanced due to resonance capture of WIMP in earth due to Fe nuclei, the sensitivity limits, assuming 100\% branching ratio for each channel, are : $\sigma_{SI} =1.02\times 10^{-44}~cm^2$  for the $\tau^{+} \tau^{-}$ channel and $\sigma_{SI} =5.36\times 10^{-44} ~cm^2$ for the $b ~\bar{b}$ channel.
\end{abstract}
\maketitle
\section{Introduction}
\label{sec:intro}

The Weakly Interacting Massive Particles (WIMP) have been proposed as one of the leading particle dark matter candidates to explain the missing non-luminous matter of the universe \cite{Zwicky:1933gu,Bertone:2004pz,Steigman:1984ac}. The WIMP with masses in the mass range of a few GeVs to tens of TeV would get gravitationally attracted to the celestial body, scatter off the nucleons in the celestial bodies and lose energy. WIMP whose velocities become less than the escape velocity of the celestial body are then trapped in the gravitational potential well of the celestial body. The trapped WIMP eventually sink to the centre of the celestial body due to gravity where their concentration increases. Subsequent annihilations of WIMP is expected to produce neutrinos in their final states. These neutrinos are expected to come with an energy spectrum in the range $[0-m_\chi]$, where $m_\chi$ is the WIMP mass. These neutrinos can be detected in the neutrino detectors, providing an indirect evidence for the existence of WIMP dark matter. Such indirect detection signals for WIMP in the sun \cite{Aartsen:2016zhm,Choi:2015ara,Albert:2016dsy} and earth \cite{Aartsen:2016fep,Albert:2016dsy} have been looked for in the currently running neutrino detectors such as IceCube \cite{Aartsen:2016zhm}, Antares \cite{Albert:2016dsy} and Super-Kamiokande (SK) \cite{Choi:2015ara}. Since none of the detectors have recorded any positive signal for WIMP annihilations in the sun and earth, they have given exclusion limits in the WIMP scattering cross-section -- WIMP mass space. 
\\


The magnetised Iron CALorimeter (ICAL) detector proposed to be built at the India-based Neutrino Observatory (INO) should be able to detect the neutrinos from WIMP annihilations if the WIMP indeed have masses in the few GeV to 100s of GeV range. In our previous work \cite{Choubey:2017vpr} we explored the prospects of indirect detection of WIMP annihilation in the sun at the ICAL detector at INO. In this work we study the prospect of indirect detection of WIMP at ICAL from their annihilations in the centre of the earth. As is well known, the WIMP annihilation cross-section can be related to the WIMP-nucleus scattering cross-section. The WIMP scattering on nucleons can proceed both via Spin Independent (SI) as well as Spin Dependent (SD) process, where the SI cross-section depends on the mass of the nucleus involved while the SD cross-section does not. Therefore, heavier target nuclei offer better sensitivity to SI cross-sections. As a result, the WIMP direct detection experiments, which look for the recoil energy of target nuclei due to WIMP scattering on them in dedicated terrestrial detectors, are more sensitivity to SI cross-sections owing to their heavier target nuclei. On the other hand, the indirect detection search for WIMP annihilation in the sun is more sensitive to the SD cross-sections since the sun mostly consists of hydrogen. The earth has heavier elements and hence can be sensitive to SI cross-sections. While the direct detection experiments continue to be several order of magnitude more sensitive for SI interactions, indirect searches provide a complementary probe to dark matter and are hence interesting. 
\\

The organisation of the paper is as follows. In Section \ref{sec:capture} we calculate the signal neutrino spectra due to WIMP annihilation in the earth. In Section \ref{sec:event} we describe the detector and the event generation procedure. Thereafter, in Section \ref{sec:background}, we describe the atmospheric background suppression scheme. In Section \ref{sec:analysis} we describe our statistical analysis, present results in Section \ref{sec:results} and finally conclude in Section \ref{sec:summary}.


\section{Neutrino flux from WIMP annihilation in the earth}
\label{sec:capture}

The number of WIMP ($N$) inside the earth as a function of time $t$ is given by the following differential equation \cite{Jungman:1995df},
\begin{equation}\label{eq:1}
 \frac{dN }{ d t}= C-C_{A} N^2-E N
 \,,
\end{equation}
where the terms on the right-hand side correspond to capture of WIMP inside the earth ($C$), annihilation in the core ($C_{A}$) and evaporation from its surface ($E$), respectively. In this work we neglect the 
effect of evaporation from the earth assuming that it is not significant for reasonably heavy WIMP \cite{Spergel:1984re,Gould:1987ju,Krauss:1985aaa,Griest:1986yu}. 
Each annihilation reduces the number of WIMP by two units and hence the rate of depletion of WIMP is twice the annihilation rate in the earth,
\begin{equation}\label{eq:4}
\Gamma_{A}=  \frac{1}{2} C_{A}N^2
\,.
\end{equation}
The quantity $C_{A}$ depends on $\langle\sigma_{A}v\rangle$ which is the total WIMP annihilation cross-section times the relative velocity of the WIMP. 
Solving Eq.~(\ref{eq:1}) for $N$, we find the annihilation rate at any given time as, 
\begin{equation}\label{eq:5}
    \Gamma_{A}=\frac{1}{2}C\tanh{^2}(t/\tau) 
    \,,
\end{equation}
where $\tau=(CC_{A})^{-1/2}$ is the time required for equilibrium to be established between the capture and annihilation of WIMP in the earth. If $t \gg \tau$, equilibrium is seen to be established and we have $\Gamma_A = C/2$. Since the capture rate $C$ depends directly on the WIMP-nucleon scattering cross-section, we get a direct relation between the annihilation rate and the WIMP-nucleon scattering cross-section. However, for the case of the earth it is seen that equilibrium has not reached and hence there is no simple proportionality between the annihilation rate and WIMP-nucleon scattering cross section. Nevertheless, $t/\tau$ can be related to the capture rate $C$ and annihilation cross-sections $\langle \sigma_{A}v \rangle$ via the following relation:
\begin{equation}\label{eq:6}
   \frac{t_{\oplus}}{\tau_{\oplus}} = 1.9\times 10^{4} \bigg (\frac{C}{s^{-1}} \bigg)^{1/2} \bigg( \frac{\langle \sigma_{A}v \rangle}{cm^3 s^{-1}} \bigg)^{1/2} \bigg( \frac{m_{\chi}}{10GeV} \bigg)^{3/4}
\end{equation}
where $t = t_{\oplus} \sim 4.5 \times 10^{9}$ years is the age of the earth. For a fixed value of $\langle \sigma_{A}v \rangle$, $C_{A}$ is constant and a direct proportionality between $\Gamma_{A}$ and 
$C$ can be established. Hence, the annihilation of WIMP in earth can be related to the 
WIMP-nucleon scattering cross section. The scattering of the WIMP could proceed via both Spin Dependent (SD) and Spin Independent (SI) processes. The SI scattering cross-section depends on the number of nucleons present in the nucleus and hence is dominant for heavy nuclei. Since the earth comprises mainly heavy nuclei, the SI WIMP-nucleon scattering is dominant and is given by \cite{Gould:1991hx,Jungman:1995df}: 
\begin{equation} 
\label{eq:2}
 C = c \left(\frac{1~\textup{GeV}}{m_\chi}\right) 
  \left( \frac{\rho_{local}}{0.3~ \textup{GeV}/\textup{cm}^{3}}  \right)  \left(\frac{270~ \textup{km}/\textup{s}}{\bar v_{\textup{local}}} \right) \sum_{i} F_{i}(m_{\chi}) 
  \sigma_{SI}^i f_{i}\phi_{i}\frac{S(m_{\chi}/m_{Ni})}{m_{Ni}/(1~{\rm GeV})}
 \,,
\end{equation}
where $c$ is a constant with value $4.8\times 10^{15}$ s$^{-1}$ for the case of earth and $m_\chi$ is the mass of WIMP.  $\rho_{local}$ and $\bar v_{\textup{local}}$ are the local DM density and velocity dispersion in the halo respectively. The summation in Eq.~(\ref{eq:2}) has to be carried out over all the nuclei in the earth where $F_i(m_\chi$) is the form-factor suppression for the capture of a WIMP of mass $m_\chi$ with the $i^{th}$ nucleus. For the $i^{th}$ nuclear species with mass $m_{N_i}$ (in GeV), $f_{i}$ and $\phi_i$ are its mass fraction and distribution in the earth respectively. $\sigma_{SI}^i$ is the cross-section for elastic scattering of the WIMP on $i^{th}$ nuclear species via SI interaction in units of $10^{-40}$ cm$^2$. For the capture of the WIMP on $i^{th}$ nuclei, $S(m_\chi / m_{N_i})$ gives the corresponding kinematic suppression factor. The cross-section for interaction of WIMP $\sigma_{SI}^i$ with $i^{th}$ nucleus can be related to WIMP-nucleon interaction cross-section $\sigma_{SI}$ by the following expression :
\begin{equation}\label{eq:7}
 \sigma^{i}_{SI} = \sigma_{SI} A^{2}_{i} \bigg( \frac{\mu_{\chi N_{i}} }{\mu_{\chi p}} \bigg)^{2}
\end{equation}
where for the $i^{th}$ nucleus: $A_i$ is the atomic number, $\mu$ is its reduced mass and $m_{N_{i}} \approx A_{i}m_{p}$, $m_{p}$ being the proton's mass. We assume here that proton mass to be equal to the neutron mass.
\begin{figure}[tbp]
\centering 
\includegraphics[width=.75\textwidth,origin=c]{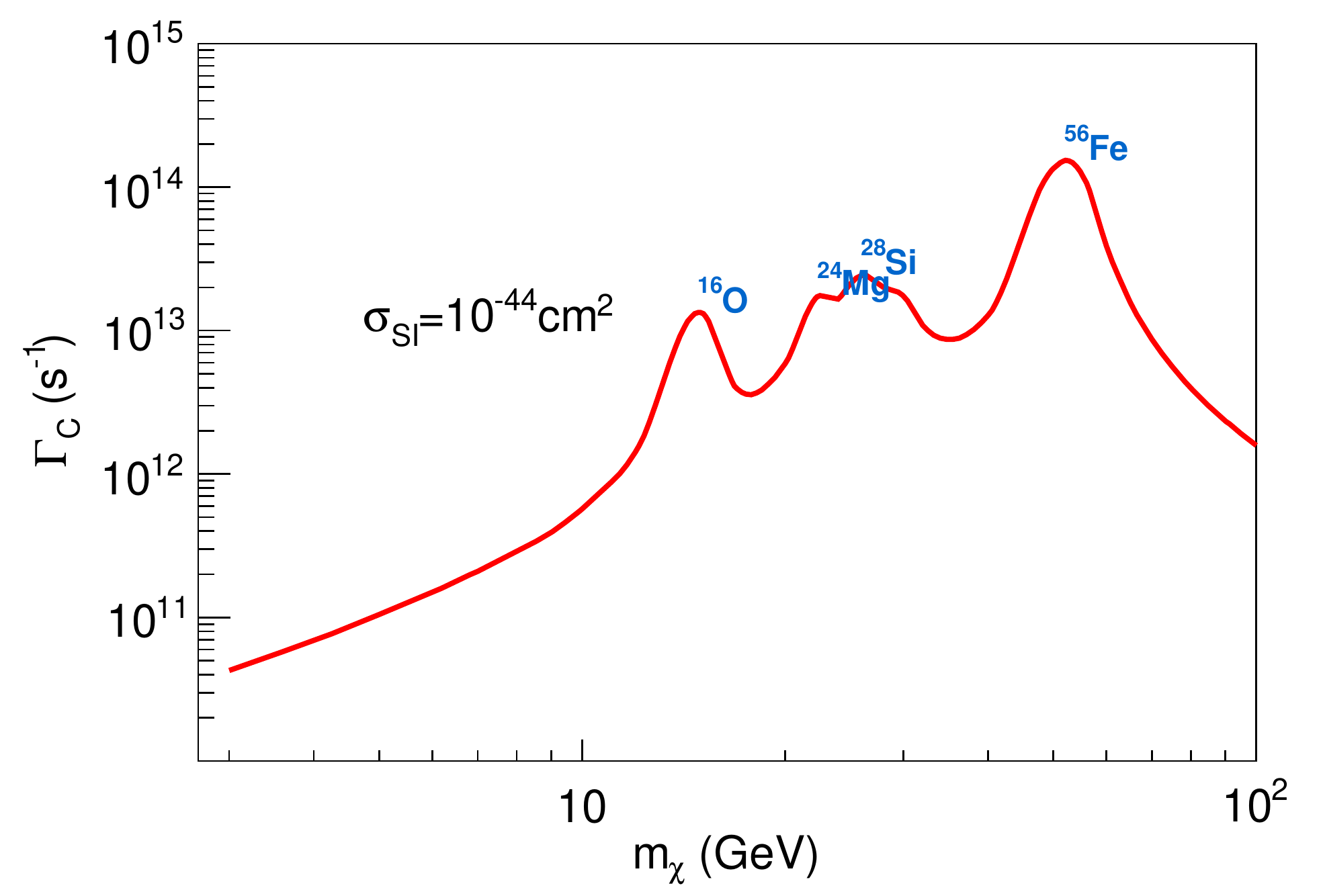}
\caption{\label{fig:1} SI capture rate for WIMP in the earth as a function of WIMP mass $m_{\chi}$. The enhanced capture rate due to resonance scattering on various nuclei is also shown.}
\end{figure}
The annihilation of WIMP produce standard model particle antiparticle pairs. Subsequently, hadronisation and/or decay of these pair products can give rise to neutrinos. 
Due to WIMP annihilation in the earth, the differential neutrino flux arising at the detector is given by:
\begin{equation}
 \frac{dN_{\nu}'}{d\Omega dt dE_{\nu}} = \frac{\Gamma_{A}}{4\pi R^{2}} \sum_{j=1} {\rm BR}_j \frac{dN_{j}}{dE_{\nu}}
 \,,
 \label{diffflux}
\end{equation}
where $\Gamma_A$ and $C$ are related as discussed above, $R$ is the distance travelled by the neutrinos between the point of creation in the earth's core and detection at the detector. $dN_j/dE_\nu$ is the differential neutrino flux for a given WIMP annihilation channel $j$ such as $W^+W^-$, $b\bar b$, $c\bar c$, $\tau^{+}\tau^{-}$ etc. The sum in Eq.~(\ref{diffflux}), which is over all possible channels $j$, has to be weighted with the branching ratio ($BR_{j}$) of the particular channel $j$. Considering a generic WIMP scenario, we take one annihilation channel at a time and assume 100\% branching ratio for each of the channels. For a specific model predicting a different $BR_{j}$, the above fluxes would be simply mixture of different channels scaled linearly.\\\\
The annihilation of WIMP into standard model particle-antiparticle pairs in the centre of the earth, followed by the propagation of the produced neutrinos up to the detector is done using the WIMPSIM \cite{Blennow:2007tw,Edsjo:2007xyz} package. WIMPSIM uses Nusigma \cite{Edsjo:2007abc} for simulation of neutrino-nucleon interactions. For the hadronisation, decay and production of 
neutrinos it uses PYTHIA \cite{Sjostrand:2006za}. We consider WIMP mass in the range $(10-100)$ GeV for both $b\bar ~b$ and $\tau^{+}\tau^{-}$ annihilation channels. The propagation of neutrinos through earth matter involves neutrino oscillations which has been incorporated in a full three flavour neutrino framework with the oscillation parameters given in Table~\ref{tab:1}. Throughout our analysis, we consider normal mass hierarchy. \\

\begin{table}[tbp]
\centering
\begin{tabular}{|lr|c|}
\hline
Paramter & Best-Fit Value\\
\hline 
$\theta_{12}$ & $34^{\circ}$ \\
$\theta_{13}$ & $9.2^{\circ}$ \\
$\theta_{23}$ & $45^{\circ}$ \\
$\delta$ & $0$ \\
$\Delta m^2_{21}$ & $7.5\times 10^{-5} \textup{eV}^2$ \\
$\Delta m^2_{31}$ & $2.4\times 10^{-3} \textup{eV}^2$\\
\hline
\end{tabular}
\caption{Oscillation parameters used in the Simulation}\label{tab:1}                                                                                                                                                
\end{table}

\begin{figure}[tbp]
\centering 
\includegraphics[width=.85\textwidth,]{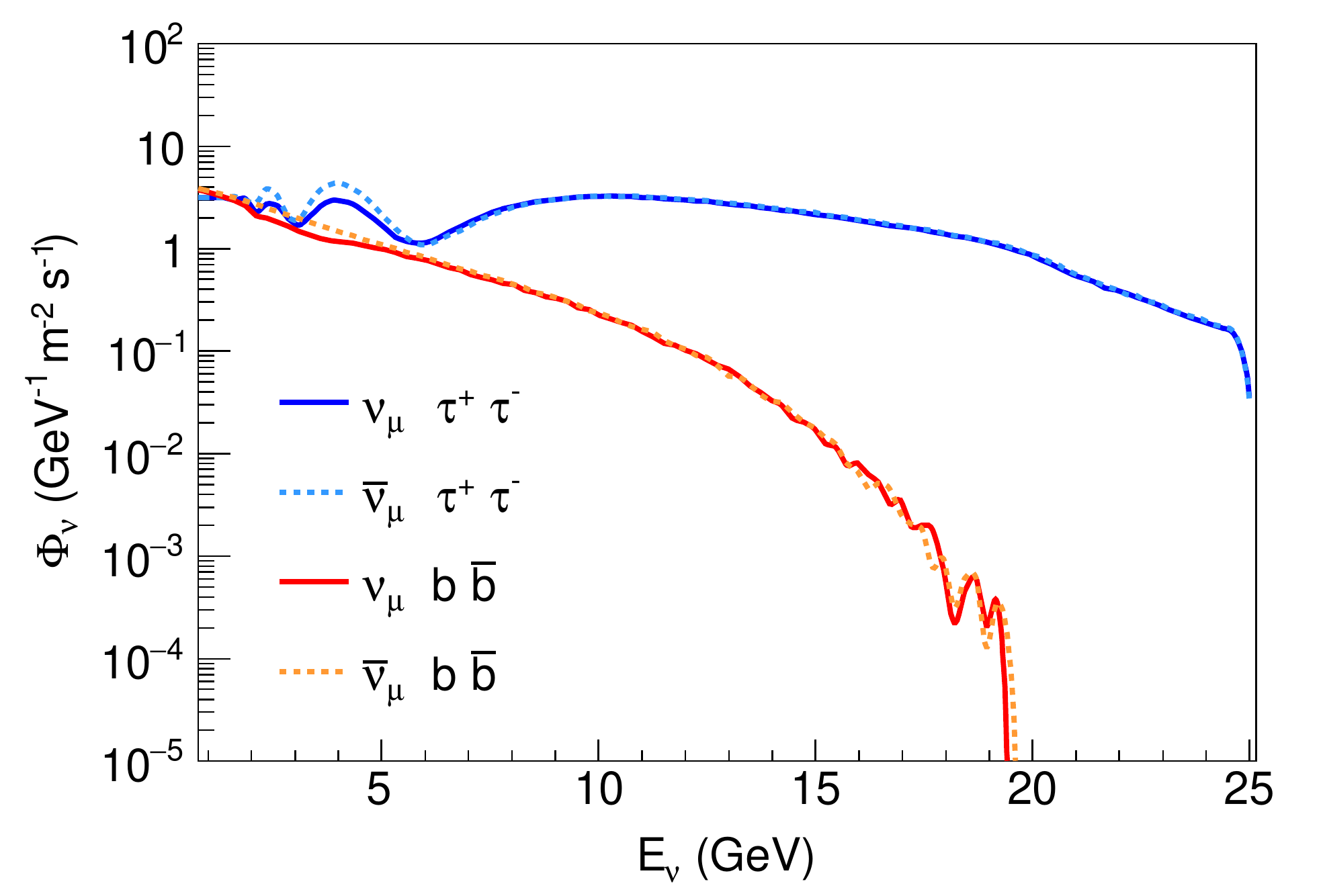}
\caption{\label{fig:2} The $\nu$ and $\bar{\nu}$ fluxes at ICAL due to annihilation of 25 GeV WIMP in the earth for channels $b \bar{b}$ and $\tau^{+} \tau^{-}$ channel 
with $\sigma_{SI}=10^{-38}$ cm$^{2}$ and $\langle \sigma_{A}v \rangle = 3 \times 10^{-26}$ cm$^3$ s$^{-1}$.}
\end{figure}
Fig.~\ref{fig:2} shows the neutrino and antineutrino fluxes (in units of GeV$^{-1}$m$^{-2}\Omega^{-1 }$s$^{-1}$) at ICAL due to WIMP annihilations in the earth. 
For each of the annihilation channels, we assume 100 \% BR.  We take $ \rho_{\textup{local}} =0.3~ \textup{GeV}/ \textup{cm}^3$ and $v_{\textup{local}}= 270 ~\textup{km}~ \textup{sec}^{-1}$ in our flux calculations. The fluxes shown in Fig.~\ref{fig:2} are for 25 GeV WIMP assuming a WIMP-nucleon scattering cross section $\sigma_{SI}=10^{-38}$ cm$^{2}$. We can see from the figure that the fluxes for 
both neutrinos as well as antineutrinos are nearly same, with antineutrino fluxes being slightly higher than the neutrino fluxes. This feature seems to hold for nearly all WIMP masses. The maximum kinematically possible energy of neutrinos produced from annihilation of WIMP is set by the WIMP mass i.e. for a WIMP mass of 25 GeV, the produced neutrinos will have energies in the range $(0-25)$ GeV.
For the $\tau^{+} \tau^{-}$ channel, the neutrino fluxes fall by about 1 order of magnitude in the above range. However, for the $b \bar{b}$ channel, the fluxes fall sharply by many orders of magnitude and well before $E_\nu = 20$ GeV it becomes negligible. Therefore, stronger indirect detection bounds are expected with neutrino fluxes from $\tau^{+} \tau^{-}$ channel in comparison to $b \bar{b}$ channel.
The fluxes arising due to annihilation channels involving the other quark-antiquark pairs are even weaker. Hence we do not consider those channels while discussing the expected sensitivity to indirect detection of dark matter in ICAL. The individual BR for a particular channel depends on a specific model considered. We, however, take a model-independent approach in this paper and quote the expected sensitivity limits for the $\tau^{+} \tau^{-}$ and $b \bar{b}$ channels with 100 \% BR each, as mentioned above. For a specific WIMP model, the flux with mixed BR for these channels will be between these extremes and hence the corresponding bounds. The Fig.~\ref{fig:2} show fluxes for benchmark values of WIMP mass and cross-sections. The fluxes for other values of $\sigma_{SI}$ can be obtained by simply scaling with the value of the cross-section. The above mentioned features of fluxes from various annihilation channels hold for all WIMP masses.
\section{Event generation at ICAL}\label{sec:event}
India-Based Neutrino Observatory (INO) is a proposed underground research facility to be built in Theni district of Tamil Nadu which is in the southern part of India. INO, among a few other experiments, will host be a 50 kt Iron CALorimeter (ICAL) detector. ICAL will have 150 layers of glass Resistive Plate Chambers (RPCs) as an active medium. Each of these RPC layers will have iron plates between them which will act as an interaction medium whereby the muon neutrinos will interact with iron and produce muons. These muons will leave long tracks. Since the iron in the detector will be magnetised, $\mu^{+}$ and $\mu^{-}$ will bend in opposite directions giving ICAL a capability of charge identification  \cite{Kumar:2017sdq}. 
ICAL, with its excellent angular resolution of muons, can be used to put limits on the WIMP annihilation from the earth, competitive with other indirect searches.\\\\
GENIE \cite{Andreopoulos:2009rq}, suitably modified for our purpose, has been used for generating neutrino events with an ICAL geometry comprising 50 kt iron mass and 150 layers of glass RPCs. We use fluxes calculated by Honda et al. \cite{PhysRevD.83.123001} for Theni site for the simulation of atmospheric neutrino background. Signal calculation is done with the fluxes as prescribed in Section \ref{sec:capture}. Events are generated for a benchmark WIMP-nucleon cross-sections of $\sigma_{SI}=10^{-40} ~cm^{2}$, assuming 100 \% BR for each of the annihilation channel, and then scaled appropriately for other WIMP-nucleon cross-sections. \\\\
We generate the signal and background events separately. Subsequently, we pass them through our reconstruction code whereby ICAL energy and angle resolutions, 
reconstruction and charge identification efficiencies are applied to get the final events at the detector. The muons, in our analysis, are binned in reconstructed energy and zenith angle bins. We perform detector simulations for ICAL geometry with Geant4 \cite{Agostinelli:2002hh} and obtain the muon reconstruction efficiency, muon charge identification efficiency, muon zenith angle resolution and muon energy resolution values. We tabulate these resolutions and efficiencies in a two dimensional table implying that the muon energy and angle resolutions are a function of both muon energy as well as muon zenith angle and are described in detail in our previous work \cite{Choubey:2017vpr}.
After incorporating the efficiencies and resolutions, the number of reconstructed $\mu^{-}$ events in the $ij^{th}$ bin are:
\begin{equation} \label{eq:3}
 N'^{th} _{ij} =  \mathcal{N} \sum_{k} \sum_{l} K^{k} _{i} (E^{k} _{T}) M^{l} _{j} (\cos{\Theta^{l} _{T}}) \left( \varepsilon_{kl} \mathcal{C}_{kl} n_{kl} (\mu^{-}) + \bar{\varepsilon_{kl}} (1- 
 \bar{\mathcal{C}_{kl}}) n_{kl} (\mu^{+}) \right) \,,
\end{equation}
where $\mathcal{N}$ is the normalisation that we require for a given exposure in ICAL. The summation in Eq.~(\ref{eq:3}) is over true muon energy and true muon zenith angle bins and are indicated by indices $k$ and $l$, respectively. $E_{T}$ and $\cos{\Theta_{T}}$ are the true (kinetic) energy and true zenith angle of the muon, respectively, whereas $E$ and $\cos{\Theta}$ are the corresponding reconstructed (kinetic) energy and zenith angle of reconstructed of $\mu^{-}$. Using the reweighting algorithm prescribed in \cite{Ghosh:2012px}, the raw $\mu^{-}$ and $\mu^{+}$ events from GENIE are folded with the three-generation oscillation probabilities and subsequently binned in terms of muon energy and zenith angles. The subsequent number of $\mu^{-}$ and $\mu^{+}$ events, in the respective $k^{th}$ true energy and $l^{th}$ true angle bin, are denoted by the quantities $n_{kl}(\mu^{-})$ and $n_{kl}(\mu^{+})$. For the $k^{th}$ energy and the $l^{th}$ zenith angle bin, the quantities $\varepsilon_{kl}$ and $\bar\varepsilon_{kl}$ are the reconstruction efficiencies of $\mu^{-}$ and $\mu^{+}$, respectively, and $\mathcal{C}_{kl}$ and $\bar{\mathcal{C}}_{kl}$ are the corresponding charge identification quantities.
\begin{figure}[tbp]
\centering
\includegraphics[width=.75\textwidth]{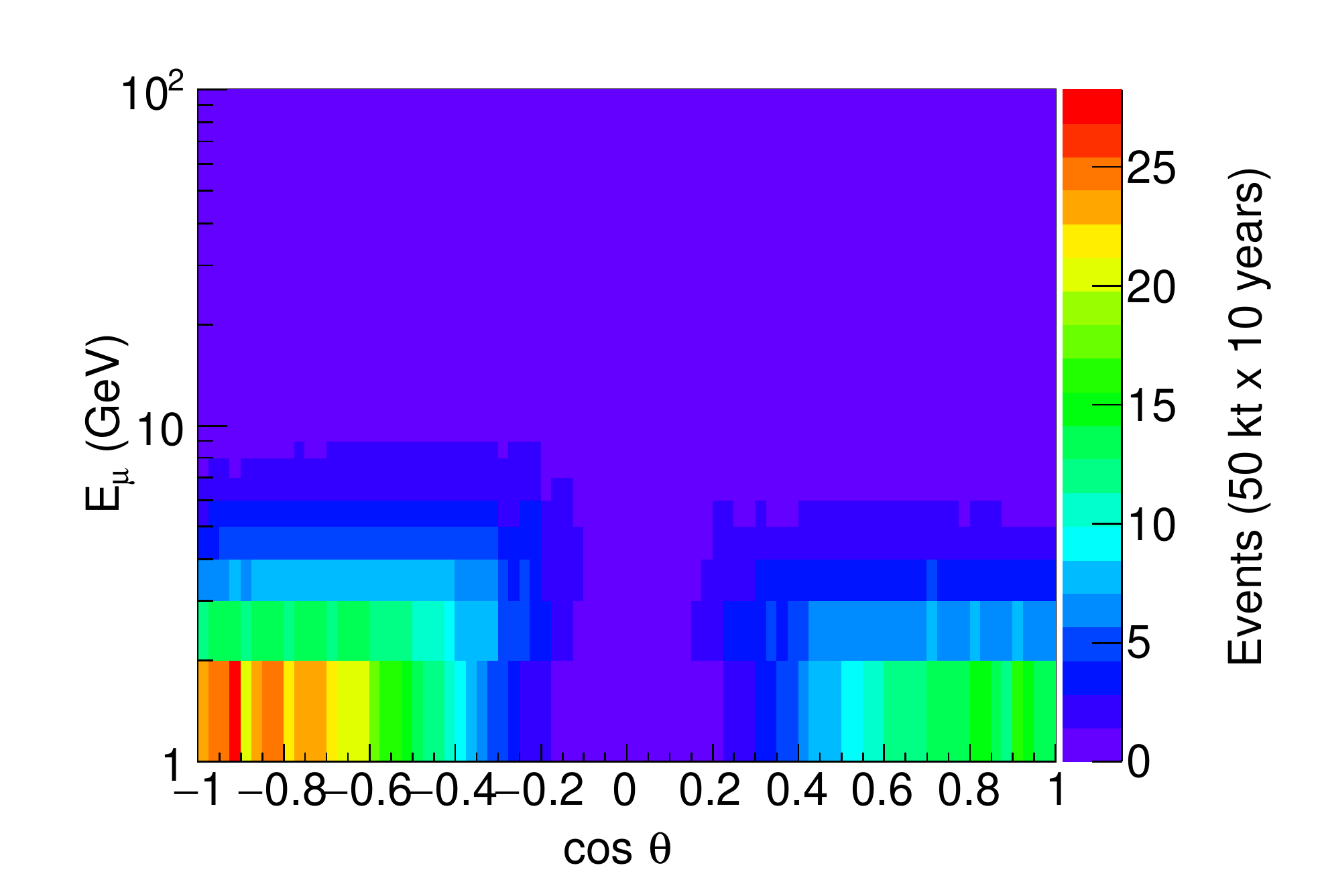}
\caption{\label{fig:3} $\mu^{-}$ event distribution at ICAL due to atmospheric neutrino background for 50 $\times$ 10 kt-years of ICAL exposure. ICAL has zero efficiency for horizontal tracks which is reflected in the bins around $\cos\theta =0$. Note that in ICAL convention $\cos\theta = 1$ represents upward going muons.}
\end{figure}
The reconstruction efficiencies as well as the charge identification efficiencies are the functions of the true muon energy $E_{T}$ and true muon zenith angle $\cos{\Theta_{T}}$. The
Gaussian resolution functions $K^{k}_{i}$ and $M^{l}_{j}$ are used to apply muon energy and angle smearing respectively and are given as:
\begin{equation} 
  K^{k} _{i}= \int _{E_{L_{i}}} ^{E_{H_{i}}} dE \frac {1} {\sqrt{2\pi}\sigma_{E}} \textup{exp} \left( -\frac {({E^{k}_{T}-E})^{2}} {2\sigma^{2}_{E}} \right)\,,
\end{equation}
\begin{equation}  f
 M^{l} _{j}(\cos_{\Theta_{T}^{l}})  = \int _{\cos{\Theta_{L_{j}}}} ^{\cos{\Theta_{H_{j}}}} d\cos{\Theta} \frac {1} {\sqrt{2\pi}\sigma_{\cos_{\Theta}}} \textup{exp} \left( -\frac 
 {({\cos{{\Theta}^{k}_{T}}-\cos{\Theta}})^{2}} {2\sigma^{2}_{\cos_{\Theta}}} \right)\,.
\end{equation}
The values of $\sigma_{E}$ and $\sigma_{\cos{\Theta}}$ are as given in appendix of \cite{Choubey:2017vpr}. Similar expressions can be written for the $\mu^+$ events and  $N'^{th}_{ij}(\mu^{+})$. All analysis in this work is done for 10 years of ICAL running.

\section{Atmospheric neutrino background suppression}
\label{sec:background}
\begin{figure}[tbp]
\centering 
\includegraphics[trim=10 10 10 10,clip,width=.45\textwidth]{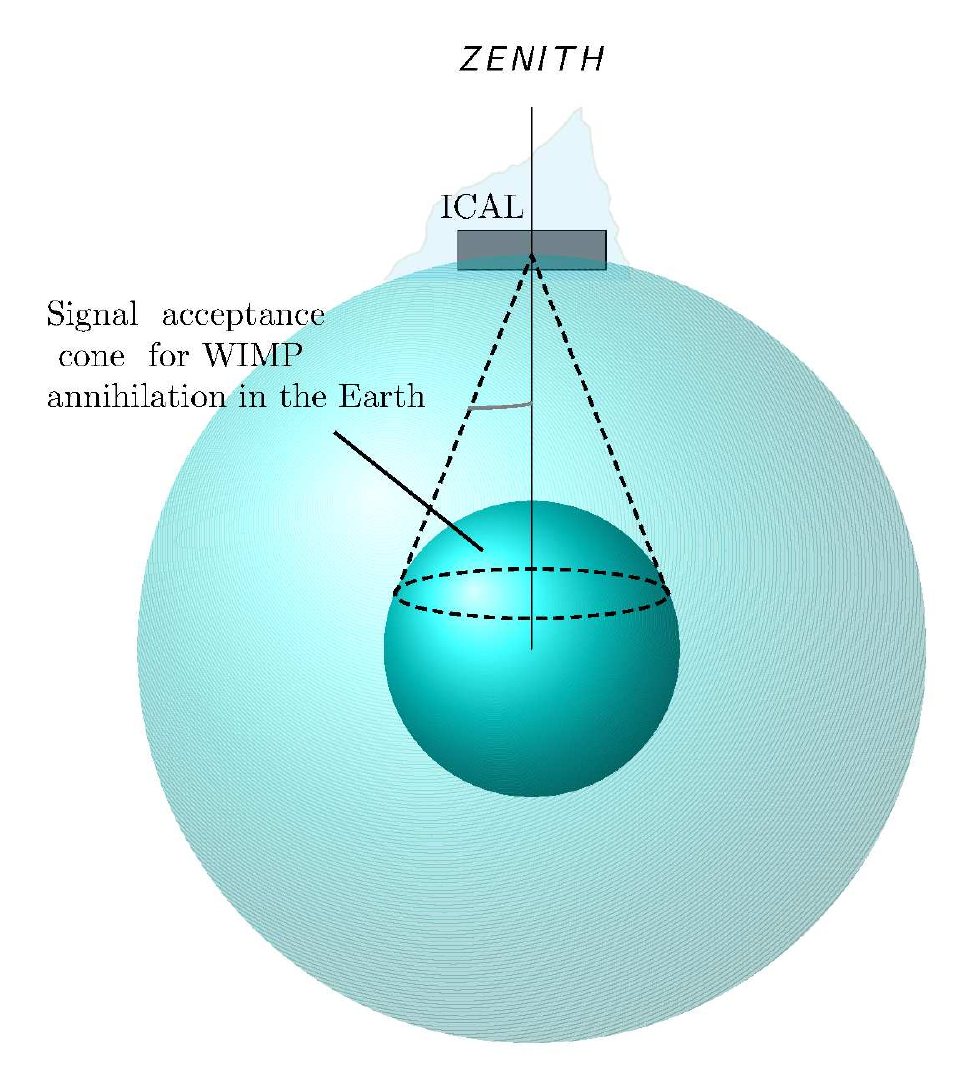} 
\caption{\label{fig:4}  The cone regions where signal from the WIMP annihilations are expected for the earth.}
\end{figure}

\begin{figure}[tbp]
\centering 
\includegraphics[width=.85\textwidth,origin=c]{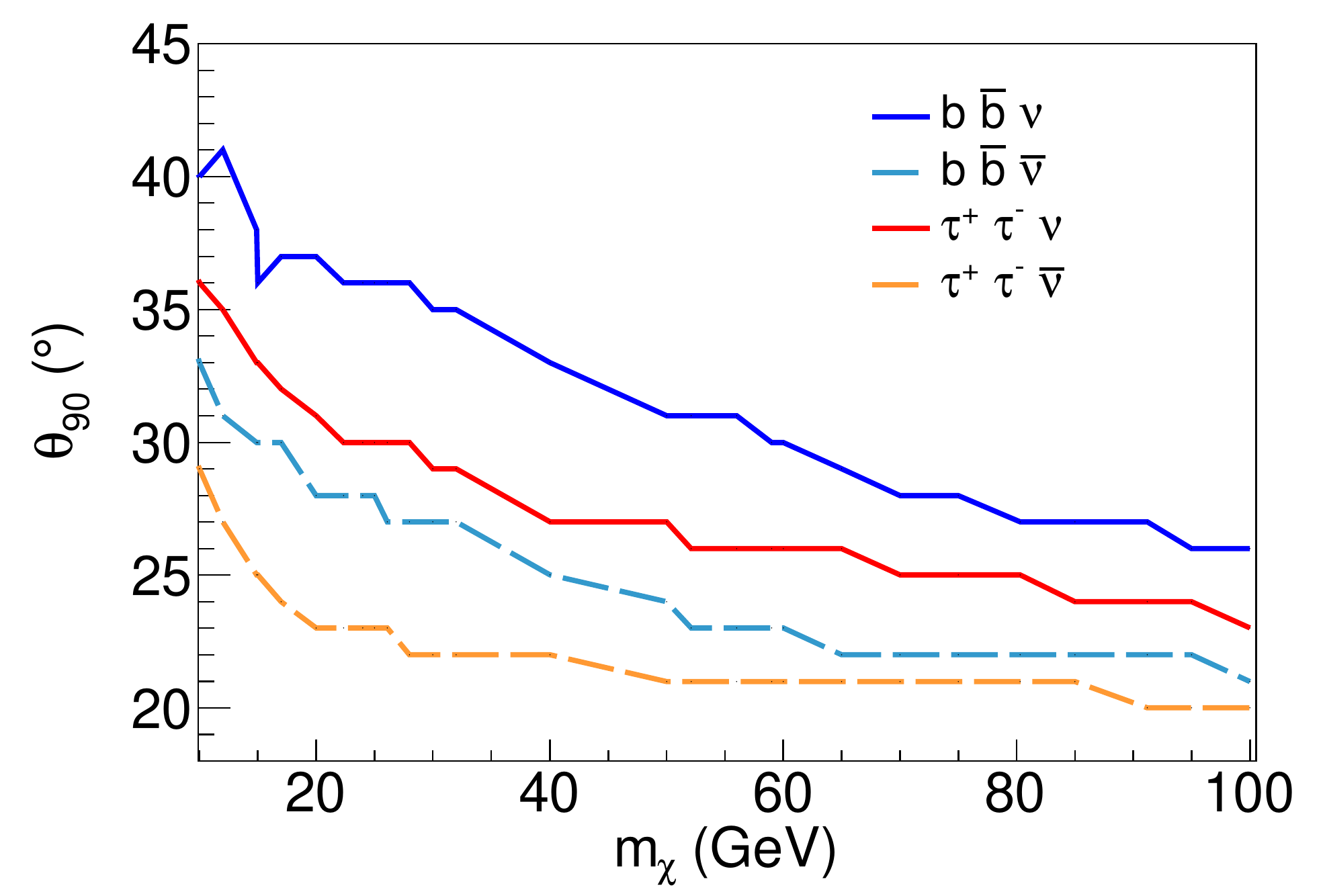}
\caption{\label{fig:5} 90 \% cone cut values obtained for the earth. This is obtained using WIMPSIM and GENIE. A cone angle is estimated such that it contains 90\% of the signal events. The solid lines correspond to neutrinos and dashed lines correspond to anti-neutrinos for each of the annihilation channel.}
\end{figure}

The major source of background to the indirect searches from WIMP annihilation in the earth is due to the atmospheric neutrinos. However, unlike the neutrinos from WIMP annihilation which come from the direction of the earth core, the atmospheric neutrinos have a distribution over all zenith and azimuth angular bins and is comparatively well studied. We can exploit this feature and use it to suppress the atmospheric background considerably. Other source of neutrinos such a geothermal neutrinos coming from the core direction are in the MeV range and hence are not relevant here. The signal neutrinos, for the case of WIMP annihilation in the earth, will come from the direction of earth core. The signal search region for the WIMP annihilation in the earth is shown in Figure~\ref{fig:4}. These neutrinos on reaching ICAL will interact with the detector iron through charge current interaction and produce charged leptons, muons being the lepton of interest for ICAL. The scattered muon will make an angle ($\theta_{\nu \mu}$) with their parent neutrino, where ($\theta_{\nu \mu}$) is a function of parent $\nu$ energy and detector medium. Due to finite detector resolution there will be smearing effects. However, we choose to work with true muon direction rather than the reconstructed muon direction at the stage of background suppression. ICAL has an excellent muon angle resolution \cite{Choubey:2017vpr} for the considered energy range and hence this choice will not affect the final results significantly.\\\\
We define $\theta_{90}$ to be the half angle of the cone that contains 90\% of the signal muons, the axis of the cone being in the direction of the earth's core. Harder channels like $\tau^{+} \tau^{-}$ with higher energy neutrinos will have a narrower $\theta_{90}$ in comparison to softer channels like $b \bar{b}$ with lower energy neutrinos which will have a broader $\theta_{90}$. Likewise, we expect that the $\nu$ spectra from annihilation of massive WIMP to have most of the associated muons in a narrower cones than neutrinos by lighter WIMP. Also, the heavier the WIMP, the closer it is to the centre of the earth and hence smaller cone opening. Using WIMPSIM and GENIE, we estimate $\theta_{90}$ for each WIMP mass and for a given annihilation channel. Figure~\ref{fig:5} shows the $\theta_{90}$ obtained for WIMP annihilation inside earth, as a function of WIMP mass ($m_{\chi}$), and for different annihilation channels.\\
\begin{figure}[tbp]
\centering 
\includegraphics[width=.75\textwidth]{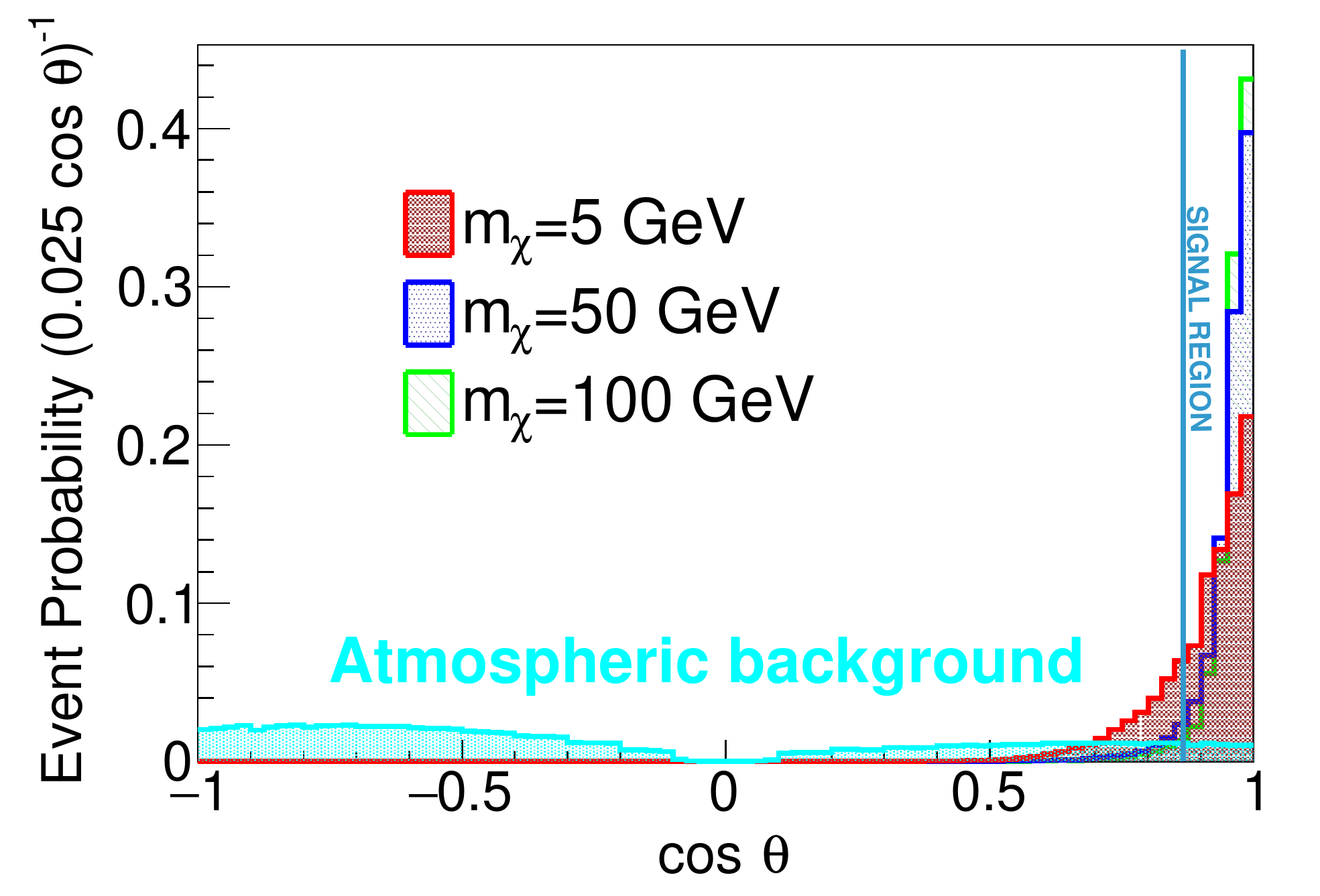}
\caption{\label{fig:8} Angular probability distribution of reconstructed $\mu^{-}$ at ICAL due to WIMP annihilations in the earth. A comparison for three WIMP masses has been shown and the region where signal is expected is marked. The probability distribution for the unsuppressed atmospheric background is also shown.}
\end{figure}

\begin{figure}[tbp]
\centering 
\includegraphics[width=.45\textwidth,]{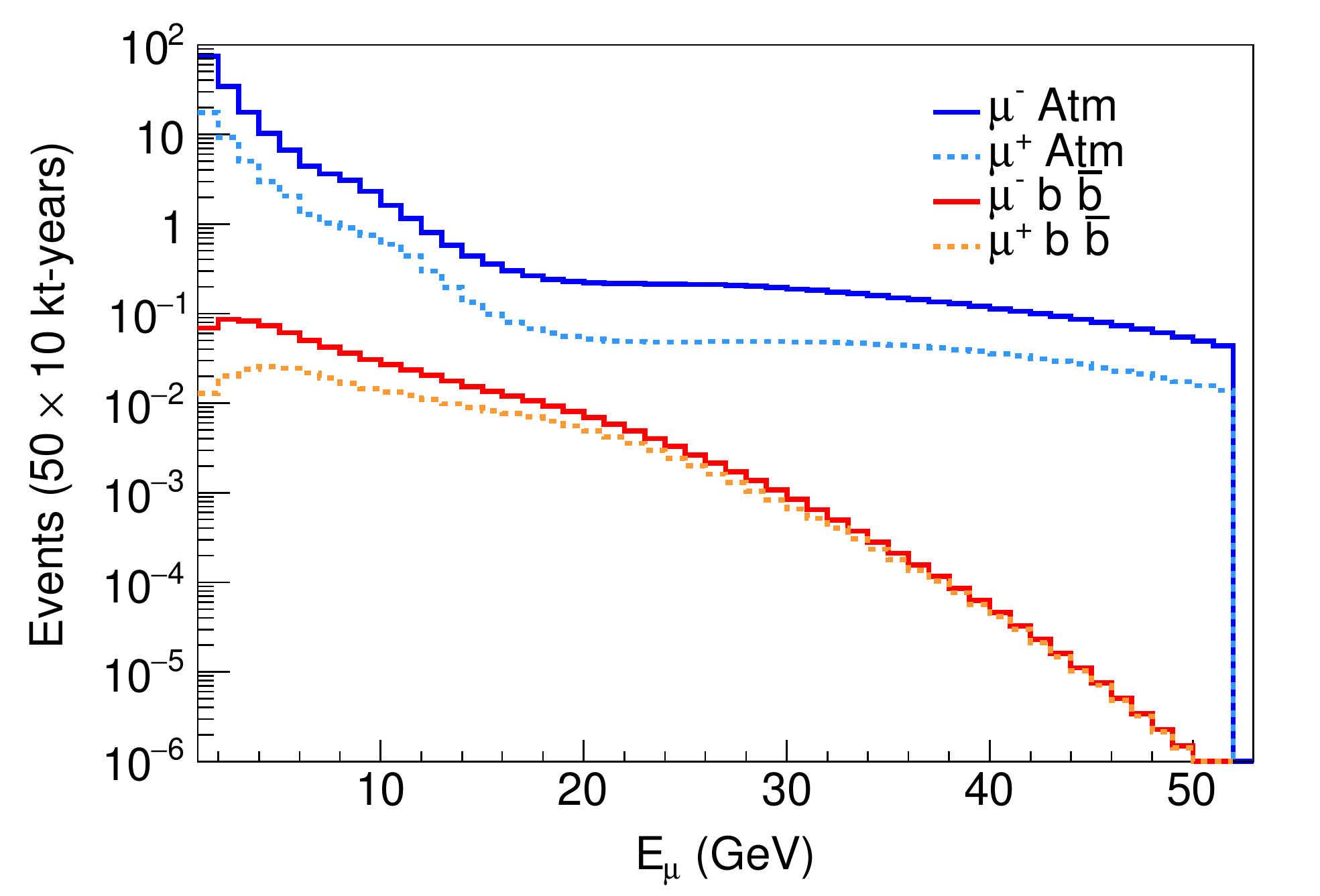}
\hfill
\includegraphics[width=.45\textwidth,origin=c]{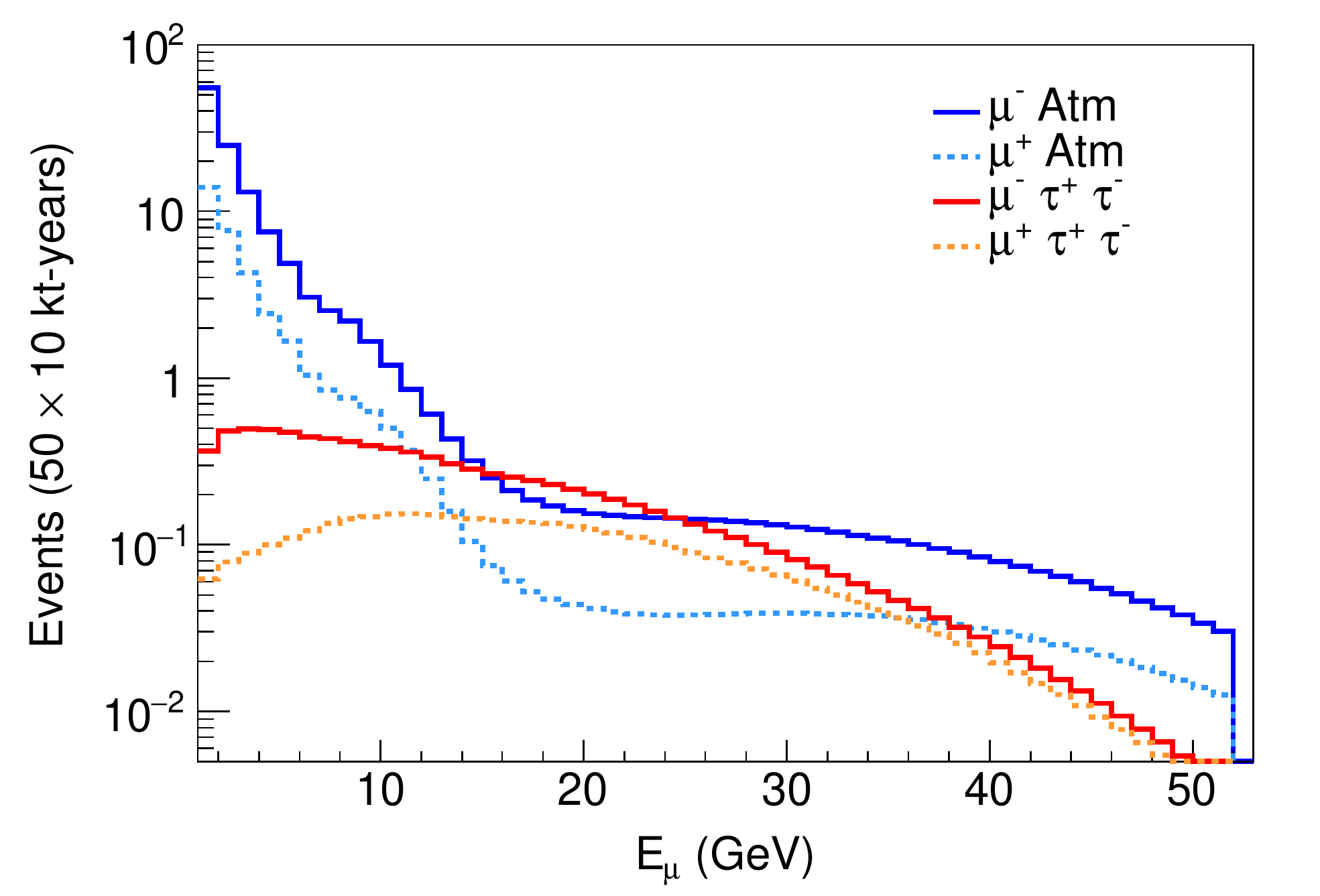}
\caption{\label{fig:9} Muon event distribution at ICAL due to atmospheric neutrinos and signal neutrinos arising out of a WIMP annihilations in the earth. The plots are for fluxes arising due
to SI capture rate. A cross-section of $\sigma_{SI} =10^{-40} cm^{2}$ has been assumed for the signal neutrinos. The left plot is for the WIMP annihilating into the channel $b \bar{b}$ while
the right plot is for the annihilation channel $\tau^{+} \tau^{-}$. A 100 \% branching ratio has been assumed for each of the channels. Also shown are the corresponding events coming from the atmospheric 
background after applying the suppression scheme as described in \ref{sec:background}. Atmospheric events have been simulated using Honda fluxes at Theni \cite{Honda:2015fha}. 
The softer channel $b \bar{b}$ has more background than the harder channel $\tau^{+} \tau^{-}$.}
\end{figure}
The atmospheric neutrino background is then suppressed as follows. For each WIMP mass and annihilation channel, we accept only those muons whose zenith angle are within $\theta_{90}$ with respect to the earth core. These atmospheric background events that fall within this cone represents an irreducible background for we can not distinguish them from the neutrinos due to WIMP annihilation in the core. After applying this suppression scheme, we fold the background events with detector resolution and efficiencies as described in the previous Section~\ref{sec:event} to obtain the final reconstructed and suppressed background events which are then used for $\chi^{2}$ analysis.\\\\
Figure~\ref{fig:8} shows the angular probability distribution of $\mu^{-}$ due to WIMP annihilations in the earth for the $\tau^{+} \tau^{-}$ channel along with the distribution of the (unsuppressed) atmospheric background muon events at ICAL. A comparison for three WIMP masses $5$, $50$ and $100$ GeV has been shown. The above probability distribution, for each of the WIMP mass, has been obtained by normalising the reconstructed $\mu^{-}$ events in each bin by total number of reconstructed $\mu^{-}$ events for that WIMP mass. It can be noted that as WIMP mass increases, the angular probability distribution peaks towards the direction of the core. This is expected because of the reasons discussed above. The signal search is carried out in the region right of the vertical line represents $\theta_{90}$. We draw a line at $\sim 30^{\circ}$ from the earth centre $(\cos\theta=1)$ just for illustration. The actual values of $\theta_{90}$ for each WIMP mass and channel is taken from Figure~\ref{fig:5} while doing the analysis. Fig.~\ref{fig:9} shows the signal events due to a 52.14 GeV WIMP annihilating through $\tau^{+} \tau^{-}$ and $b \bar{b}$ channels. Also shown are corresponding suppressed atmospheric background events.

\section{The statistical analysis}
\label{sec:analysis}
We estimate the 90 \% C.L. sensitivity limits on SI WIMP-nucleon cross-sections through a $\chi^{2}$ analysis. In our analysis, we generate prospective data at ICAL comprising atmospheric background events only. To this `data', we fit our hypothesis in which we consider combined events predicated at ICAL due to WIMP annihilation in the earth and atmospheric neutrino background. This choice is consistent with `no WIMP scenario' in the data and hence the limits calculated are the expected exclusion limits in the WIMP mass - WIMP SI cross-section plane from 10 years of running of ICAL.

We combine the $\mu^{+}$ and $\mu^{-}$ events while performing $\chi^2$ analysis.
A $\chi^2$ function is defined as 
\begin{equation}
 \chi^2 = \chi^2(\mu^-) +  \chi^2(\mu^+)
\end{equation}
where 
\begin{equation}
\chi^2(\mu^\pm) = \min_{\xi^\pm_k}\sum_{i=1}^{N_i}\sum_{j=1}^{N_j}
\bigg [ 2\bigg(N_{ij}^{\rm th}(\mu^\pm) - N_{ij}^{\rm ex}(\mu^\pm)\bigg ) + 
2N_{ij}^{\rm ex}(\mu^\pm)\ln\bigg(\frac{N_{ij}^{\rm ex}(\mu^\pm)}{N_{ij}^{\rm th}(\mu^\pm)}
\bigg ) \bigg]
+ \sum_{k=1}^l {\xi^\pm_k}^2
\,,
\end{equation}
\begin{equation}
N_{ij}^{\rm th}(\mu^\pm) = {N'}_{ij}^{\rm th}(\mu^\pm)\bigg(1+\sum_{k=1}^l \pi_{ij}^k{\xi^\pm_k}\bigg)  
+{\cal O}({\xi^\pm_k}^2)\,,
\end{equation}
where ${N'}_{ij}^{\rm th}(\mu^\pm)$ are the $\mu^\pm$ events that we `predict' and $N_{ij}^{\rm ex}(\mu^\pm)$ are the events `observed' at ICAL. For $k^{th}$ systematic uncertainty we have associated $\pi_{ij}^k$ correction factors with $\xi^\pm_k$ being the corresponding pull parameters. Similar to our previous analysis \cite{Choubey:2015xha}, we include 5 systematic errors as follows. We take 20~\% error on neutrino flux normalisation and 10~\% error on neutrino-nucleon cross-section. On the zenith angle distribution of atmospheric neutrino fluxes, we include a 5~\% uncorrelated error and 5~\% tilt error. Finally, we take a 5~\% overall error to account for detector systematics. 
Minimisation over the pull parameters gives the individual contributions from $\mu^-$ and $\mu^+$ data samples. Thereafter, we add them up and calculate the $\chi^2$ for a given set of WIMP mass and WIMP-nucleon cross-sections.
\section{Results}
\label{sec:results}
\begin{figure}[tbp]
\centering 
\includegraphics[width=.65\textwidth]{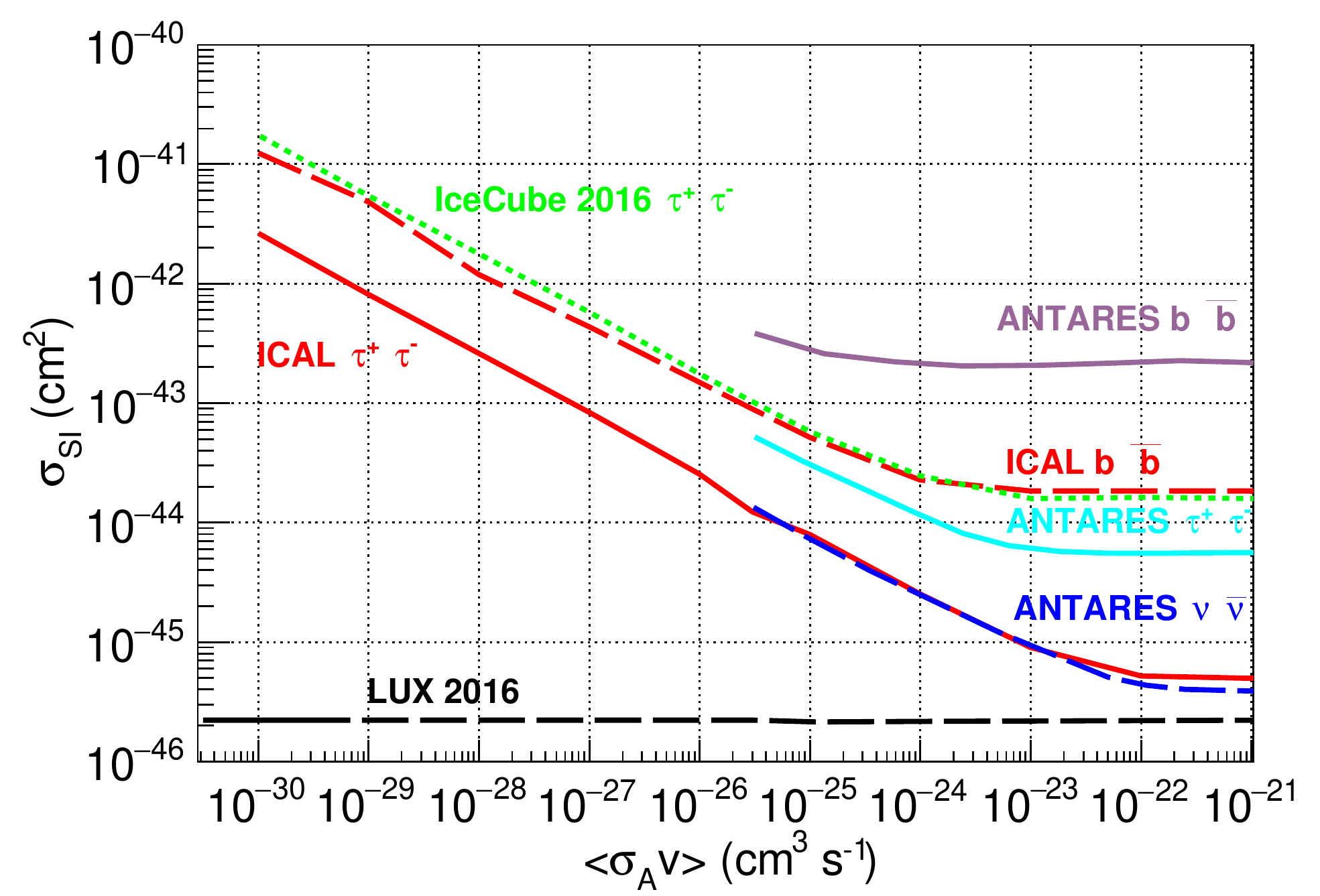}
\caption{\label{fig:10} The expected 90 \% C.L. sensitivity limits on $\sigma_{SI}$ for ICAL as a function of annihilation cross-section $\langle \sigma_{A}v\rangle$ for a $50 GeV$ WIMP annihilating through  $\tau^{+}$ and $\tau^{-}$ (red solid) and $b ~\bar{b}$ (red dashed). Also shown are upper limits at 90 \% C.L obtained by various experiments IceCube \cite{Aartsen:2016fep} $\tau^{+}$ and $\tau^{-}$ (green), ANTARES\cite{Albert:2016dsy} $\tau^{+}$ and $\tau^{-}$ (cyan), $b ~\bar{b}$ (light purple), $\nu_{\mu}\bar\nu_{\mu}$ (blue) and LUX\cite{Akerib:2016vxi} (black-dashed) have been shown for comparison. 
For ICAL, systematics have been included.}
\end{figure}

\begin{figure}[tbp]
\centering 
\includegraphics[width=.85\textwidth]{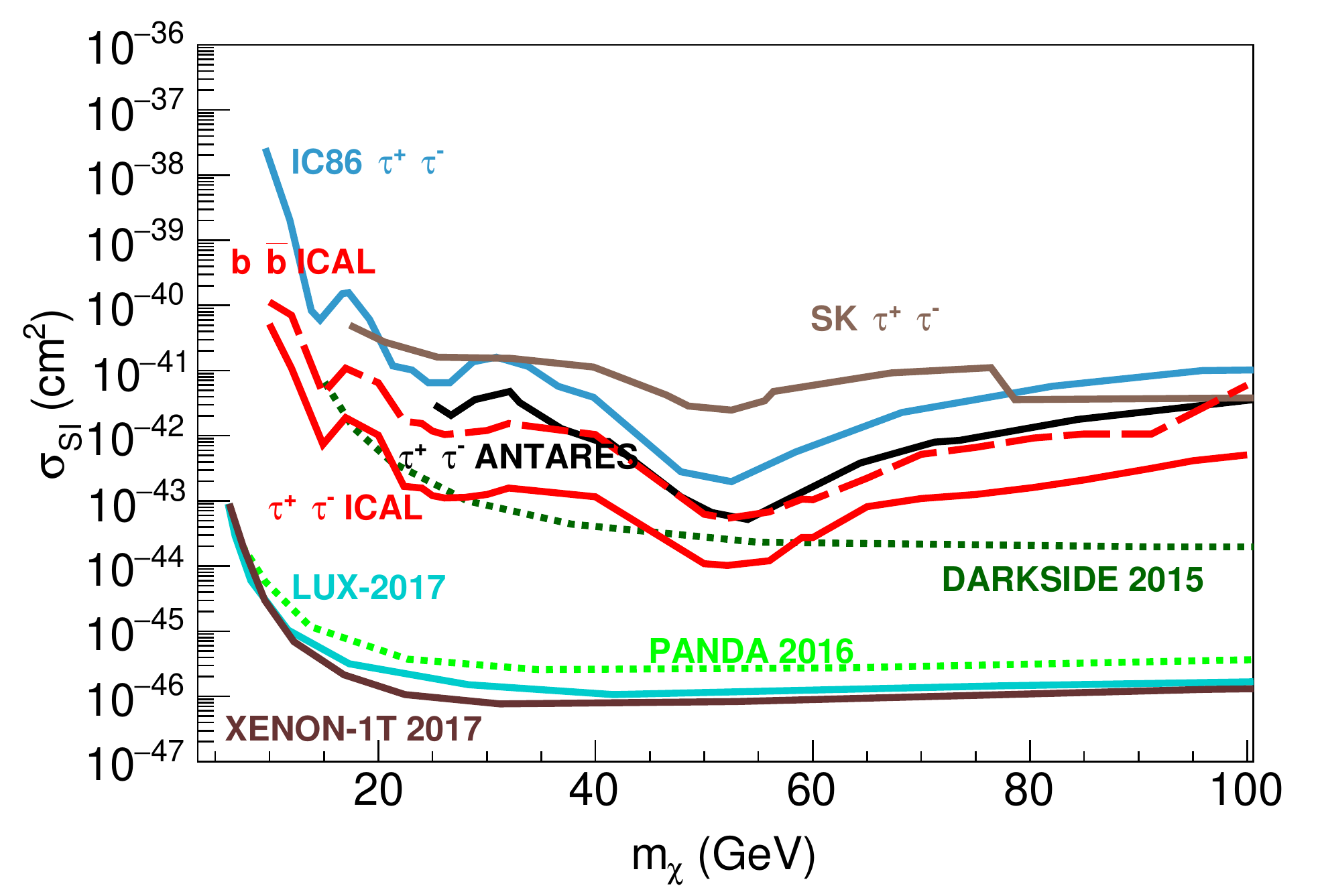}
\caption{\label{fig:15} The expected 90\% C.L. sensitivity limits on $\sigma_{SI}$ as a function of WIMP mass, assuming a WIMP annihilation cross section 
$\langle\sigma_{A}v\rangle= 3\times 10^{-26} cm^{3}s^{-1}$. The displayed sensitivity limits are for the local dark matter density $\rho =0.3 GeV cm^{-3}$. Among the indirect detection experiments, 
ICAL provides the most stringent bound. The dip around 50 GeV in the limits obtained for Earth WIMP annihilation is a prominent feature in all experiments and is due to 
resonant capture of WIMPs on Fe. ICAL 90\% C.L. sensitivity limits for $\tau^{+}$ and $\tau^{-}$ (red solid) and $b ~\bar{b}$ (red dashed) for WIMP annihilation in the Earth are shown; 90\% C.L. upper limits from SK\cite{Desai:2004pq} for $\tau^{+}$ and $\tau^{-}$ (brown), IceCube\cite{Aartsen:2016fep} $\tau^{+}$ and $\tau^{-}$ (blue) and ANTARES \cite{Albert:2016dsy} 
$\tau^{+}$ and $\tau^{-}$ (black). Also, shown are the limits obtained from DARKSIDE \cite{Agnes:2014bvk} (dark green dotted), LUX (cyan), XENON-1T \cite{Aprile:2017iyp} (brown) and PANDA\cite{Tan:2016zwf} (light green dotted).}
\end{figure} 

\begin{figure}[tbp]
\centering 
\includegraphics[width=.85\textwidth]{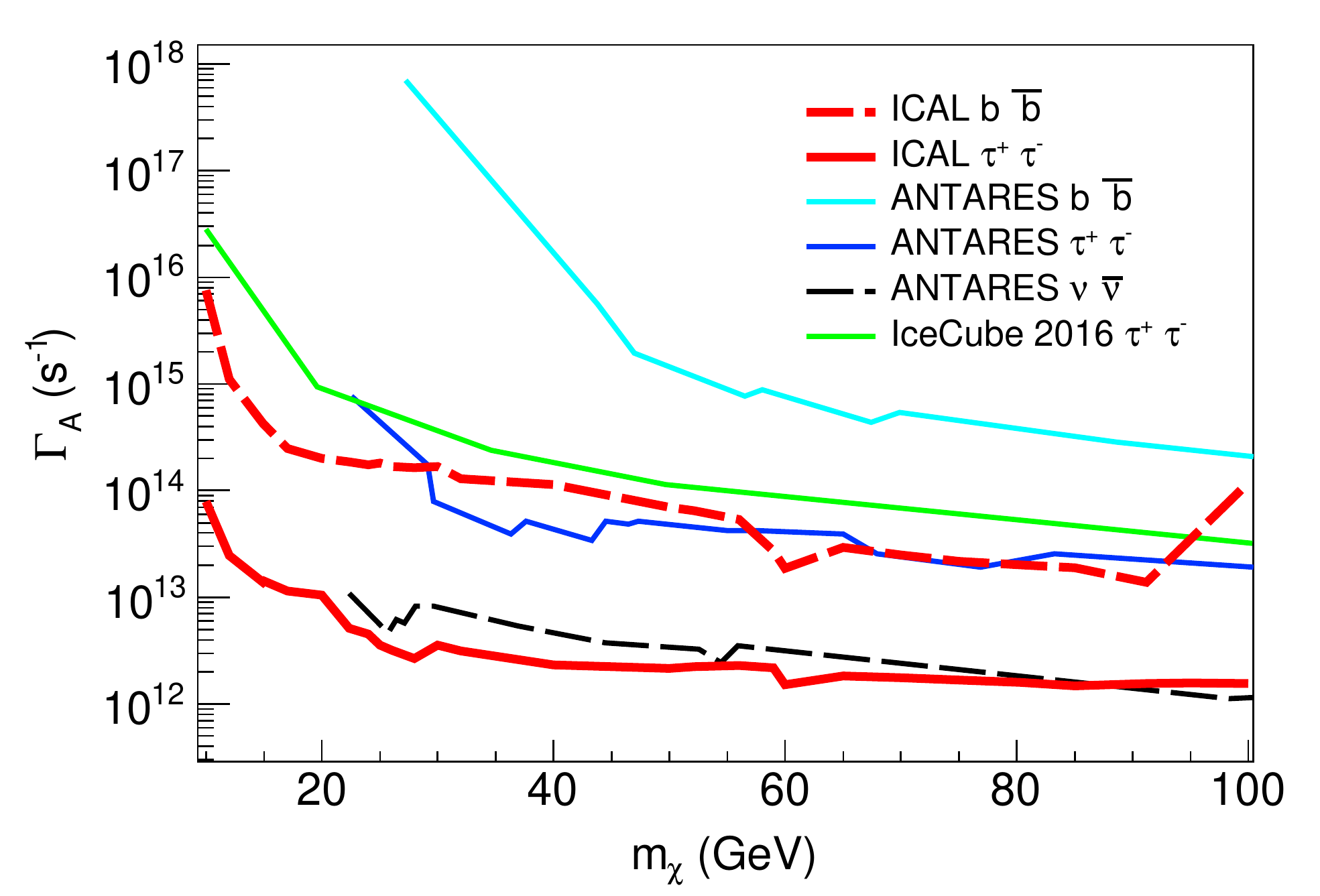}
\caption{\label{fig:11} The expected 90\% C.L. sensitivity limits on the annihilation rate ($\Gamma_{A}$) as function of WIMP mass ($m_{\chi}$) due to WIMPS annihilating into $\tau^{+}$ $\tau^{-}$ 
(red solid) and $b ~\bar{b}$ (red dashed) with 100\% branching ratio each. The limits are for 10 years of ICAL running. For comparison limits from various other experiments have been shown: 
ANATARES\cite{Albert:2016dsy} $\tau^{+}$ $\tau^{-}$ (blue), $b ~\bar{b}$ (cyan), $\nu_{\mu}\bar\nu_{\mu}$ (black) , and IceCube\cite{Aartsen:2016fep} $\tau^{+}$ and $\tau^{-}$ for 
$m_{\chi} <50 GeV$ and $W^{+} W^{-}$ for $m_{\chi} >50 GeV$ (green). We show these limits for a fixed $\langle\sigma_{A}v\rangle = 3 \times 10^{-26} cm^{3} s^{-1}$.}
\end{figure}

\begin{figure}[tbp]
\centering 
\includegraphics[width=.85\textwidth]{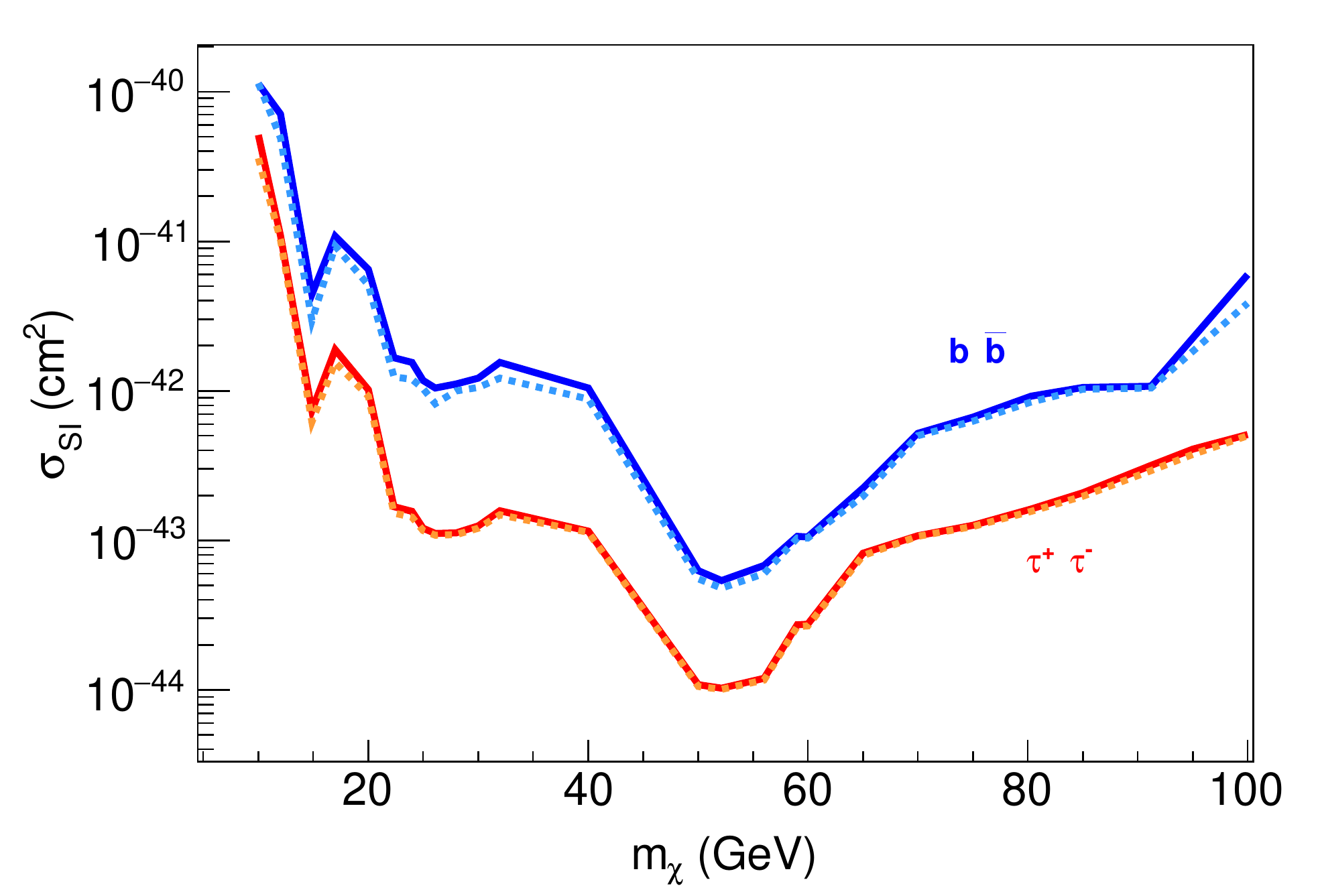}
\caption{\label{fig:12} The expected 90\% C.L. sensitivity limits for WIMP annihilation for different masses and two annihilation channels for WIMP annihilation in the earth are shown. The solid lines are the sensitivity limits calculated using detector systematics as described in Sec~\ref{sec:analysis}. The corresponding dotted lines are without systematics. The expected sensitivity limits for $\tau^{+}$ and $\tau^{-}$ are shown in red (with systematics) and orange-dotted lines (without systematics) and for $b ~\bar{b}$ in blue (with systematics) and azure-dotted lines (without systematics). The effect of systematics, as we expect, is to worsen the limits as expected.}
\end{figure}

We present the main results in this section. As mentioned in Section~\ref{sec:capture}, for the case of WIMP annihilation in the earth, the spin-independent WIMP nucleon cross sections $\sigma_{SI}$ and annihilation rate $\Gamma_{A}$ are related by Eq.~(\ref{eq:4}-\ref{eq:6}). The capture rate and the annihilation rate are not in equilibrium, and hence the annihilation rate depends on the $\sigma_{SI}$ as well as on the annihilation cross section $\langle\sigma_{A}v\rangle$. Figure~\ref{fig:10} shows our expected sensitivity upper limits at 90 \% C.L. in the $\sigma_{SI}$ - $\langle\sigma_{A}v\rangle$ plane for a WIMP mass of $50$ GeV. As shown in Figure~\ref{fig:1}, the capture rate for the WIMP masses closer to iron mass is greatly enhanced and hence we expect stronger bounds. It is evident that for a $50$ GeV WIMP, ICAL seems to put a stronger bound on $\sigma_{SI}$ for a given $\langle\sigma_{A}v\rangle$ for a given annihilation channel such as $\tau^{+}$ $\tau^{-}$. For the obvious reasons, described in earlier sections, the harder channel $\tau^{+}$ and $\tau^{-}$ gives a stronger limit in comparison to the softer channel $b ~\bar{b}$.\\\\
In Figure~\ref{fig:15} we present the expected 90 \% C.L. sensitivity limits on the WIMP-nucleon SI interaction cross-section as a function of WIMP mass for $500$ kt-years of ICAL exposure and compare it with the exclusion limits obtained from various other direct and indirect detection experiments. The direct detection experiment XENON-1T \cite{Aprile:2017iyp} gives the most stringent bound till date. Bounds from indirect searches are, in general, weaker in comparison to direct detection experiments. However, among the neutrino detectors, ICAL seems to give the most stringent bound for the chosen WIMP mass range. For the WIMP masses close to iron mass, there is a resonant capture and hence enhance event rates resulting in a stronger bound. \\\\
For calculating sensitivity limits on annihilation rate and $\sigma_{SI}$ in case of the earth, we assume an annihilation cross-section $\langle\sigma_{A}v\rangle =3 \times 10^{-26} cm^{3}s^{-1}$. As discussed in Section~\ref{sec:capture}, we have a relation between the annihilation rate $\Gamma_{A}$ and SI WIMP-nucleon cross-section $\sigma_{SI}$. Using the $\chi^{2}$ analysis described in Section~\ref{sec:analysis}, we derive limits on the $\sigma_{SI}$ as a function of WIMP mass $m_{\chi}$. Using Eq.~(\ref{eq:4}-\ref{eq:6}) from Section~\ref{sec:capture}, we transported these sensitivity limits from $\sigma_{SI} -m_{\chi} $ plane to $\Gamma_{A} -m_{\chi} $ plane. Figure~\ref{fig:11} shows the expected sensitivity limits at 90 \% C.L. calculated on the WIMP annihilation rate for annihilation in the earth through channels $\tau^{+}$ and $\tau^{-}$ and $b ~\bar{b}$. Results from other experiments are also shown for comparison\footnote{For SK latest preliminary results please see \cite{Frankiewicz:2017trk}.}. Again, we can see that for chosen WIMP mass range, ICAL presents a stronger bound on the WIMP annihilation rate for a given channel. Again, $\tau^{+}$ $\tau^{-}$ bounds are stronger than that of $b ~\bar{b}$. \\\\
Finally, in Figure~\ref{fig:12} we present the effect of systematic uncertainties on the expected 90 \% C.L. sensitivity limits on the WIMP-nucleon SI interaction cross-section. The dashed lines are the limits calculated while taking only statistical uncertainties. As expected, the effect of systematic uncertainties is to worsen the limits as can be seen from the figure.

\section{Summary}
\label{sec:summary}
The analysis presented in this work is a part of ongoing studies to probe the physics potential of the upcoming ICAL detector. Neutrinos arising out of WIMP annihilations in the earth could be used to probe dark matter signatures. Such searches would be complementary to direct searches for WIMP. We presented a study of prospects of detecting muon events at ICAL arising due to WIMP annihilation in the earth for $ \tau^+ \tau^- $ and $ b~\bar b$ annihilation channels. Employing an effective atmospheric background suppression scheme, the expected 90~\% C.L. sensitivity limits obtained for SI WIMP-nucleon cross-section for the case for the earth is better than any other indirect detection experiment. 
\acknowledgements
We gratefully acknowledge INO collaboration for the support. We sincerely thank Amol Dighe for intensive discussions. We acknowledge the HRI cluster computing facility (http://cluster.hri.res.in). The authors would like to thank the Department of Atomic Energy (DAE) Neutrino Project under the XII plan of Harish-Chandra Research Institute. This project has received funding from the European Union's Horizon 2020 research and innovation programme InvisiblesPlus RISE under the Marie Sklodowska-Curie grant agreement No 690575. This project has received funding from the European Union's Horizon 2020 research and innovation programme Elusives ITN under the Marie Sklodowska- Curie grant agreement No 674896. This project has received funding from FONDECYT 3170845 (Chile).
\bibliographystyle{JHEP}
\bibliography{ref}
\end{document}